\documentclass[lettersize,journal]{IEEEtran}
\usepackage{graphicx}
\usepackage{graphics}
\usepackage{colortbl}
\usepackage{multicol}
\usepackage{diagbox}
\usepackage{lipsum}
\usepackage{color}
\usepackage[cmex10]{amsmath}
\usepackage{amsthm}
\usepackage{amssymb}
\usepackage{mathrsfs}
\usepackage{mathtools}
\usepackage{amsbsy}
\usepackage[colorlinks=true,bookmarks=false,citecolor=blue,urlcolor=blue]{hyperref}
\usepackage[dvipsnames]{xcolor}\usepackage{mathrsfs}
\usepackage{setspace}
\usepackage{float}
\usepackage{graphicx}
\usepackage{pstool}
\usepackage{epstopdf}
\usepackage{lettrine}
\usepackage{psfrag}
\usepackage{newclude}
\usepackage{dsfont}
\usepackage[normalem]{ulem}
\usepackage{latexsym}
\usepackage{algpseudocode}
\usepackage{algorithm,algpseudocode}
\usepackage{algorithmicx}
\usepackage{gensymb}
\usepackage{marginnote}
\usepackage{multirow}
\usepackage{amsmath}
\usepackage{amsmath,bm}
\usepackage{wasysym}
\usepackage{mathrsfs}
\usepackage{amsmath,cite,amsfonts,amssymb,color}
\usepackage[T1]{fontenc}
\usepackage{graphicx,psfrag}
\usepackage{pstool}
\usepackage{algorithm}
\usepackage{algpseudocode}
\usepackage{comment}
\usepackage{tabularx}
\usepackage{caption}
\usepackage{csquotes}
\usepackage{float}
\usepackage{lettrine}
\usepackage{makecell}
\usepackage{textcomp, gensymb}
\usepackage{relsize}
\usepackage{balance}
\usepackage{lipsum}

\ifCLASSOPTIONcompsoc
\usepackage[caption=false,font=normalsize,labelfon
t=sf,textfont=sf]{subfig}
\else
\usepackage[caption=false,font=footnotesize]{subfig}
\fi
\allowdisplaybreaks

\graphicspath{{figures/}}
\allowdisplaybreaks

\newcommand{\RNum}[1]{\uppercase\expandafter{\romannumeral #1\relax}}

\usepackage{mathtools}

\usepackage{comment}
\begin{document}
\title{Outage-Constrained Sum Secrecy Rate Maximization for STAR-RIS with Energy-Harvesting Eavesdroppers}

\author{\IEEEauthorblockN{Zahra Rostamikafaki, Francois Chan,and  Claude D'Amours  \thanks{Zahra Rostamikafaki, Francois Chan,and  Claude D'Amours are with the Department of Electrical Engineering and Computer Science, University of Ottawa, Ottawa, ON K1N 6N5, Canada (E-mails: \{zrost034, cdamours,  fchan2\}@uottawa.ca).
Francois Chan is also with the Department of Electrical and Computer Engineering, Royal Military College of Canada, Kingston, ON K7K 7B4, Canada (e-mail: chan-f@rmc.ca).}
}}

%\markboth{IEEE COMMUNICATIONS LETTERS,~Vol.~XX, No.~XX, 2023}%


\maketitle

\begin{abstract}
This article proposes a novel strategy for enhancing secure wireless communication through the use of a simultaneously transmitting and reflecting reconfigurable intelligent surface (STAR-RIS) in a multiple-input single-output system. In the presence of energy-harvesting eavesdroppers, the study aims to maximize the secrecy rate while adhering to strict energy harvesting constraints. By dynamically manipulating the wireless environment with the STAR-RIS, the research examines the balance between harvested energy and secrecy rate under two key protocols: energy splitting and mode selection. The study addresses both imperfect and perfect channel state information (CSI) and formulates a complex non-convex optimization problem, which is solved using a penalty concave convex procedure combined with an alternating optimization algorithm. The method optimizes beamforming and STAR-RIS transmission and reflection coefficients to achieve a optimal balance between secure communication and energy harvesting constraints. Numerical simulations show that the proposed approach is effective, even with imperfect CSI, and outperforms conventional RIS methods in terms of robust security and energy performance.
\end{abstract}


\section{Introduction}
\IEEEPARstart{T}{he} rapid pace at which wireless communication technology has advanced has resulted in a significant increase in data transmission, thereby raising crucial concerns regarding physical layer security, especially in the context of the future sixth-generation (6G) networks \cite{wu2021intelligent,renzo2019smart}. Intelligent reflecting surfaces (RISs) have emerged as a promising technology to improve network performance\cite{mu2021capacity}, and the recent development of simultaneous transmission and reflection reconfigurable intelligent surfaces (STAR-RISs) has further broadened their potential.
STAR-RISs offer a unique capability to transmit and reflect incident signals simultaneously, providing 360-degree coverage and dynamic control over signal propagation through adjustable transmission and reflection coefficients (TaRCs)\cite{liu2021star}. In secure communication systems, especially in the age of the internet of things, protecting information from eavesdroppers while maintaining efficient energy use has become critical. Traditional beamforming approaches frequently fail in situations when the channel responses of legitimate users and eavesdroppers are highly correlated \cite{li2021outage}. However, because radio frequency (RF) signals convey both information and energy, this idea can be applied to wireless power transfer, in which information receivers decode information and energy receivers capture energy from the RF signals. This dual capacity is incorporated in simultaneous wireless information and power transfer, which allows for information and energy transmission\cite{clerckx2018fundamentals}. The implementation of STAR-RIS technology increases flexibility by improving quality-of-service for authorized users while reducing information leakage to eavesdroppers and optimizing energy harvesting.

Prior research on RIS transmission with security often assumes perfect channel state information (CSI) for eavesdroppers, which is unrealistic, especially with multiple eavesdroppers. Despite existing channel estimation methods, secure transmission must account for CSI uncertainty\cite{li2020probabilistic}. 

Several studies on STAR-RIS assisted secure wireless networks have been published. Using a multi-antenna base station and STAR-RIS to optimize energy harvesting and information freshness in a wireless sensor network is explored in \cite{kavianinia2023age}. It develops scheduling and optimization techniques to minimize age of information while ensuring energy requirements are met, demonstrating improved performance compared to conventional RIS.
In \cite{niu2021simultaneous}, the use of STAR-RIS for boosting security in a multiple-input single-output (MISO) network by optimizing beamforming and TaRCs across three protocols is investigated. Simulations confirm STAR-RIS's effectiveness.  A STAR-RIS secure wireless system with energy-harvesting eavesdroppers, optimizing both secrecy and energy harvesting via TaRCs is studied in \cite{kavianinia2023star}. The non-convex problem is reformulated into a convex one, with results demonstrating the advantages of optimizing STAR-RIS and \cite{zhang2022secrecy} addresses secure transmission in STAR-RIS-assisted uplink non-orthogonal multiple access systems. It optimizes secrecy for both full and statistical eavesdropping CSI scenarios using adaptive and constant-rate wiretap codes. An alternating hybrid beamforming algorithm is used for joint optimization of beamforming, power, and STAR-RIS settings. Results show the scheme’s effectiveness and provide insights on STAR-RIS deployment. 

This paper investigates the application of energy harvesting eavesdroppers and secrecy rate measurements in  wireless communication systems to fulfill the need for energy harvesting demand and secure information transmission. Moreover, conventional RIS implementations have limitations in terms of half-space coverage. To overcome this issue, this research presents STAR-RIS, which addresses the limitations of conventional RIS by providing 360-degree wireless coverage for two legal users while also accommodating energy-harvesting eavesdroppers in the context of imperfect CSI, which has not received significant attention in the existing literature. Therefore, this paper focuses on maximizing the sum secrecy rate in a MISO wiretap network, while also ensuring that energy-harvesting eavesdroppers receive the minimum required energy. We achieve this by optimizing both the transmit power and TaRCs of STAR-RIS by applying the penalty concave convex procedure (PCCP) based on an alternating optimization (AO) method, and taking into account the challenges posed by imperfect CSI for various STAR-RIS protocol. Given that perfect CSI is often impractical in real-world scenarios, we compare the performance of our proposed system under both imperfect and perfect CSI conditions.

%To address the non-convexity of the optimization problem, characterized by a non-concave objective function and highly-coupled variables, we employ advanced optimization techniques, including a path-following method and an alternating optimization approach. Our results demonstrate that STAR-RIS significantly enhances the sum secrecy rate while fulfilling the energy harvesting requirements of eavesdroppers, even under imperfect CSI. Additionally, our study highlights the superiority of STAR-RIS over conventional RIS in secure communication, offering valuable insights for the design of next-generation wireless networks that prioritize both security and energy efficiency.

\indent\textit{Organization:} 
Section II outlines the proposed system model, STAR-RIS configuration, sum secrecy rate formulation, and secrecy outage probability. Section III delves into the optimization problem and presents the proposed solution. Section IV offers numerical results, while Section V concludes the paper.

\indent\textit{Notations:} 
$\big|\boldsymbol{x}\big|$ defines the Euclidean norm of the complex-valued vector $\boldsymbol{x}$, and the real component of $x$ is $\mathfrak{R}\{x\}$. The probability that the random variable $x$ is less than $a$ is indicated by Pr$\{x<a\}$. A zero-mean and unit-variance complex symmetric Gaussian random variable is $x \sim \mathcal{C}\mathcal{N} (0,1)$. $\boldsymbol{x}^H$ represents the vector's conjugate transpose. Furthermore, $\textrm{diag}(\boldsymbol{x})$ represents a diagonal matrix whose off-diagonal elements are zero and whose diagonal components are made up of the elements of vector $\boldsymbol{x}$.

\section{System Model}
\begin{figure}[t!]
    \centering
    \pstool[scale=0.7]{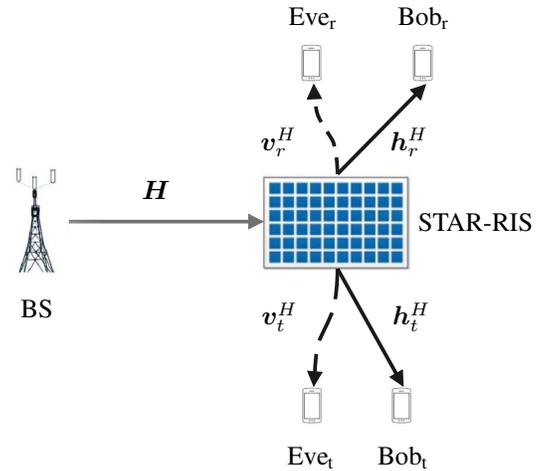}{
    \psfrag{A}{\hspace{-0.25cm}$\text{Bob}_\text{r}$}
    \psfrag{B}{\hspace{-0.25cm}$\text{Eve}_\text{r}$}
    \psfrag{C}{\hspace{-0.25cm}$\text{Eve}_\text{t}$}
    \psfrag{D}{\hspace{-0.25cm}$\text{Bob}_\text{t}$}
    \psfrag{E}{\hspace{-0.25cm}$\text{STAR-RIS}$}
    \psfrag{F}{\hspace{-0.25cm}$\boldsymbol{H}$}
    \psfrag{G}{\hspace{-0.25cm}$\boldsymbol{h}_r^H$}
    \psfrag{H}{\hspace{-0.25cm}$\boldsymbol{v}_r^H$}
    \psfrag{I}{\hspace{-0.25cm}$\boldsymbol{h}_t^H$}
    \psfrag{J}{\hspace{-0.25cm}$\boldsymbol{v}_t^H$}
    \psfrag{K}{\hspace{-0.15cm}$\text{BS}$}
    }
    \caption{Secure STAR-RIS system model.}
    \label{fig:1}
\end{figure}

 Fig. \ref{fig:1} depicts a downlink STAR-RIS-aided secrecy MISO communication system. The system is made up of a BS with ($N_t$ antennas), a STAR-RIS with $M$  simultaneously reflecting and transmitting elements, two single-antenna legitimate users ($\text{Bob}_t$ and $\text{Bob}_r$), and two single-antenna energy harvesting eavesdroppers ($\text{Eve}_t$ and $\text{Eve}_r$). $\text{Bob}_t$ and $\text{Eve}_t$  are located in the STAR-RIS' transmission region, while $\text{Bob}_r$ and $\text{Eve}_r$ are in its reflection coverage region.
 The Bob in the $k$-th area is referred to as the $\text{Bob}_k$, $k \in \{t, r\}$, and the same goes for Eves. According to the given assumption, the direct connections between the BS and both users and Eves are obstructed due to the unfavorable propagation conditions. As a result, communication can only be established through the STAR-RIS. All wireless channels, which encompass the main channel and the wiretapping channels, are subject to Rayleigh fading and are disturbed by additive white Gaussian noise (AWGN). To enhance link performance, a STAR-RIS is implemented to optimize the sum secrecy rate while fulfilling harvested energy at the eavesdropping receivers. It is assumed that the BS has statistical CSI of the channel used by STAR-RIS to communicate with Eves owing to passive and unauthorised properties of eavesdroppers.
\subsection{Signal Model of STAR-RIS Protocols}
Assume STAR-RIS includes $M$ elements, where $M$ indicates the size of STAR-RIS. The transmitted and reflected signals on a given element of STAR-RIS, $m \in \mathcal{M}  \triangleq \{ 1, 2, ..., M\}$, are represented by $t_m =\sqrt{\alpha_m^t}e^{j\phi_m^t}u_m$ and $r_m = \sqrt{\alpha_m^r}e^{j\phi_m^r}u_m$, respectively, where the incident signal on the $m$-th element is denoted by $u_m$. $\sqrt{\alpha_m^t} \in \left [0,1\right] $,
$\sqrt{\alpha_m^r} \in \left [0,1\right]$ and
$\phi_m^t \in \left [0,2\pi\right] $, $\phi_m^r \in \left [0,2\pi\right]$, denote the amplitude and phase shift of the $m$-th element, respectively. According to \cite{xu2021star}, an ideal STAR-RIS with tunable surface electric and magnetic impedance, can select $\phi_m^t$ and $\phi_m^r$  independently. In addition, $\sqrt{\alpha_m^t}$ and $\sqrt{\alpha_m^r}$ must meet the energy conservation condition $\alpha_m^t + \alpha_m^r = 1$, for any $m \in \mathcal{M}$ \cite{liu2021star}. There are two potential protocols for STAR-RIS operation, and we will briefly outline current STAR-RIS operating schemes below:
\subsubsection{The Energy Splitting Protocol}
This protocol enables all STAR-RIS components to function in transmission and reflection modes at the same time.
The TaRCs are denoted as $\boldsymbol{\Phi}_t^{ES}=\textrm{diag}(\sqrt{\alpha_1^t}e^{j\phi_1^t}, \dots , \sqrt{\alpha_M^t}e^{j\phi_M^t})$, and $\boldsymbol{\Phi}_r^{ES}=\textrm{diag}(\sqrt{\alpha_1^r}e^{j\phi_1^r}, \dots , \sqrt{\alpha_M^r}e^{j\phi_M^r})$, where $\alpha_m^t,\alpha_m^r \in \left[0,1\right]$, $\alpha_m^t + \alpha_m^r =1$, and $\phi_m^t,\phi_m^r \in \left[0,2\pi\right), \forall m\in\mathcal{M}$. 
The energy splitting (ES) protocol allows for a high level of system design flexibility since both the transmitting and reflecting coefficients of each element can be tuned.
\subsubsection{The Mode Selection Protocol}
This protocol divides STAR-RIS into two parts: one with $M_t$ elements in the transmitting mode and another with $M_r$ elements in the reflecting mode,  where $M_t + M_r = M$. Therefore, the TaRCs are represented by $\boldsymbol{\Phi}_t^{MS}=\textrm{diag}(\sqrt{\alpha_1^t}e^{j\phi_1^t}, \dots , \sqrt{\alpha_M^t}e^{j\phi_M^t})$, and $\boldsymbol{\Phi}_r^{MS}=\textrm{diag}(\sqrt{\alpha_1^r}e^{j\phi_1^r}, \dots , \sqrt{\alpha_M^r}e^{j\phi_M^r})$, where $\alpha_m^t,\alpha_m^r \in \{0,1\}, \alpha_m^t + \alpha_m^r =1$, and $\phi_m^t,\phi_m^r \in \left[0,2\pi\right)$.
The mode selection (MS) protocol is more practical than ES as it allows for "on-off" operations. The term "on-off" in this context refers to mode selection, not the functioning of pin diodes incorporated in the RIS \cite{zhang2022intelligent}.
\subsection{Signal Transmissions and Receptions}
BS delivers separate precoded signals $\boldsymbol{x}= \sum_{k} \boldsymbol{\mathit{w}}_k u_k $ to each Bob at the same frequency, where $k \in \{t,r\}$. Let $u_k$ represent the signal for $\text{Bob}_k$, with $u_k \sim \mathcal{C}\mathcal{N} (0,1)$ indicates the information symbol of $\text{Bob}_k$, and $\boldsymbol{w}_k \in \mathbb{C}^{N_t \times 1}$ specifies the BS transmit beamforming vector. The signal received at $\text{Bob}_k$ is as follows
\begin{equation}
   y_{b,k} = \boldsymbol{h}_k^H\boldsymbol{\Phi}_k\boldsymbol{H}(\sum_{k}\boldsymbol{w}_ku_k) + n_{b,k}, \forall k \in \{t,r\}\label{eq:1},
\end{equation}
where $\boldsymbol{h}_k^{H} \in \mathbb{C}^{1 \times M}$, $\boldsymbol{H} \in \mathbb{C}^{M \times Nt}$, and $\boldsymbol{\Phi}_k \in \mathbb{C}^{M \times M}$ represent the channels between  the STAR-RIS and $\text{Bob}_k$, between the BS and the STAR-RIS, and STAR-RIS coefficient for $\text{Bob}_k$, respectively.   $n_{b,k} \sim \mathcal{C}\mathcal{N} (0,\sigma_{b,k}^2)$ denotes AWGN at $\text{Bob}_k$.
Also, in eavesdropping terminals, the received signal at $\text{Eve}_k$ is given by
\begin{equation}
    y_{e,k} = \boldsymbol{v}_k^H\boldsymbol{\Phi}_k\boldsymbol{H}(\sum_{k}\boldsymbol{w}_ku_k) + n_{e,k}, \forall k \in \{t,r\}\label{eq:2},
\end{equation}
where $\boldsymbol{v}_k^{H} \in \mathbb{C}^{1 \times M}$ is the channel between the STAR-RIS and $\text{Eve}_k$, and $n_{e,k} \sim \mathcal{C}\mathcal{N} (0,\sigma_{e,k}^2)$ represents AWGN at $\text{Eve}_k$.\\
According to \eqref{eq:1} and \eqref{eq:2}, the signal-to-interference-plus-noise ratio (SINR) at the  $\text{Bob}_k$ and $\text{Eve}_k$ are respectively given by
\begin{equation}
    \gamma_{b,k} = \frac{\big| \boldsymbol{\theta}_k^H \boldsymbol{F}_k \boldsymbol{w}_k\big|^2}{\big| \boldsymbol{\theta}_k^H \boldsymbol{F}_k \boldsymbol{w}_{\acute{k}}\big|^2 + \sigma_{b,k}^2},\label{eq:3}
\end{equation}
\begin{equation}
    \gamma_{e,k} = \frac{\big| \boldsymbol{\theta}_k^H \boldsymbol{V}_k \boldsymbol{w}_k\big|^2}{\big| \boldsymbol{\theta}_k^H \boldsymbol{V}_k \boldsymbol{w}_{\acute{k}}\big|^2 + \sigma_{e,k}^2},\label{eq:4}
\end{equation}
where, considering $\boldsymbol{\theta}_k = \text{diag}(\boldsymbol{\Phi}_k)$ yields $\boldsymbol{h}_k^H\boldsymbol{\Phi}_k\boldsymbol{H} = \boldsymbol{\theta}_k^H\boldsymbol{F}_k$ and $\boldsymbol{v}_k^H\boldsymbol{\Phi}_k\boldsymbol{H} = \boldsymbol{\theta}_k^H\boldsymbol{V}_k$, where $\boldsymbol{F}_k = \text{diag}(\boldsymbol{h}_k^H)\boldsymbol{H}$ and $\boldsymbol{V}_k = \text{diag}(\boldsymbol{v}_k^H)\boldsymbol{H}$, respectively. Furthermore, $k = t$, $\acute{k} = r$, and vice versa. 

We assume $\text{Eve}_k$ eavesdrops on $\text{Bob}_k$'s information and harvests energy in the same coverage region of STAR-RIS for $\text{Bob}_k$. Then,
based on \eqref{eq:3} and \eqref{eq:4}, the sum secrecy rate (SSR) for the $\text{Bob}_k$ is computed as
\begin{equation}
    \text{SSR} = \sum_{k}[\log_2{(1 + \gamma_{b,k})} - \log_2{(1 + \gamma_{e,k})}]^+.\label{eq:5}
\end{equation}

The BS's knowledge of $\text{Eve}_k$ is imprecise due to the lack of a perfect CSI $\boldsymbol{v}_k$ \cite{mu2017secure}. As a result, a secrecy outage occurs at the BS when $\text{Eve}_k$'s channel capacity exceeds the $\text{Bob}_k$'s redundancy rate, denoted by $S_k$. Thus, the secrecy outage probability (SOP) induced by $\text{Eve}_k$ is given by \cite{li2020massive,li2021outage}
\begin{equation}
    p_{so}^k = \text{Pr}\bigg\{S_k < \log_2{(1 + \gamma_{e,k})} \bigg\}, \forall k \in \{t,r\}. \label{eq:6}
\end{equation}
In the non-collaborative eavesdropping framework, where Eves do not share their observations or outputs, the achievable SSR is given by \cite{wang2016secrecy}
\begin{equation}
    \text{SSR} =  \sum_{k}[\log_2{(1 + \gamma_{b,k})} - S_k]^+,
\end{equation}
representing the minimum of the sum secrecy rates attained by the BS in presence of Eves and $\big[x\big]^+ \triangleq \max\{x, 0 \}$.

Due to energy harvesting capability of Eves, by disregarding noise power relative to the received signal in \eqref{eq:2}, the harvested energy at $\text{Eve}_k$ is given by
\begin{equation}
    E_{e,k} = \eta \bigg(\big| \boldsymbol{\theta}_k^H \boldsymbol{V}_k \boldsymbol{w}_k\big|^2 + \big| \boldsymbol{\theta}_k^H \boldsymbol{V}_k \boldsymbol{w}_{\acute{k}}\big|^2\bigg),
\end{equation}
where $0 \le \eta \le 1$ represents the energy harvesting efficiency. For the remainder of the work, we assume $\eta = 1$.
\section{Optimization Problem Formulation}
Our objective is to develop a highly secure transmission system with respect to the SOP constraint and transmit power budget which optimizes the achievable SSR by joint optimization of the TaRCs and beamforming. Additionally, we aim to ensure that the minimum required harvesting energy is satisfied at Eves. Thus, for the ES protocol, the problem of maximizing the achievable SSR is presented as follows:
\begin{subequations}
\label{P1}
\begin{align}
\textrm{ES:}\max_{\boldsymbol{w}_k, \boldsymbol{\theta}_k} 
 &  \quad \sum_{k}[\log_2{(1 + \gamma_{b,k})} - S_k]^+,\\
\textrm{s.t.} \quad & \sum_{k} {\big| \boldsymbol{w}_k \big|}^2 \le P_{\text{max}},\label{eq:9b}\\
\quad & p_{so}^k \le \delta, \forall k, \label{eq:9c} \\
\quad & E_{e,k} \ge E_{\text{min}}, \forall k, \label{eq:9d}\\
\quad & \big[\boldsymbol{\theta}_k \big]_m = \sqrt{\alpha_m^k}e^{j\phi_m^k}, \forall k,m,  \label{eq:9e} \\ 
\quad & \alpha_m^k \in \big[0,1 \big], \phi_m^k \in \big[0,2\pi \big), \forall  k,m,\label{eq:9f} \\
\quad & \alpha_m^t + \alpha_m^r = 1,\forall  m \in \mathcal{M}, \label{eq:9g}
\end{align}
\end{subequations}
where $P_{\text{max}}$ at the BS is the maximum transmit power, and \eqref{eq:9b} denotes the maximum transmission power  constraint, a predetermined upper bound denoting the highest acceptable SOP is represented by $\delta \in (0,1)$, and SOP constraint given by \eqref{eq:9c}, \eqref{eq:9d} indicates that the harvested energy must be greater equal than the required energy $E_{\text{min}}$ at $\text{Eve}_k$, and STAR-RIS TaRCs configuration constraints represented by \eqref{eq:9e}-\eqref{eq:9g}.

The non-convex optimization problem and the probabilistic constraint present difficulties in solving \eqref{P1}. In order to overcome these difficulties, we will address the non-convex optimization problem after obtaining a closed-form formula for the SOP constraint.
\subsection{Addressing the SOP Probabilistic Constraint}
To begin with, we focus on addressing the shortcoming of \eqref{P1}'s SOP constraint. The SOP is the probability of an outage resulting from Rayleigh fading, as indicated in \eqref{eq:6}. By making use of the exponential distribution of the received signal power\cite{stuber2001principles,li2021outage}, we can derive a closed-form expression for SOP. The proof of this theorem can be found in Appendix A.\\
\textbf{Theorem 1.} \textit{The closed-form equation for SOP is provided by $p_{so}^k = \exp{\Big(-\frac{(\acute{\boldsymbol{\Psi}}_k^H\acute{\boldsymbol{\Psi}}_k + \sigma_{e,k}^2)(2^{S_k} - 1)}{\boldsymbol{\Psi}_k^H\boldsymbol{\Psi}_k} \Big)}, \forall k$, where 
$\boldsymbol{\Psi}_k = \boldsymbol{\Phi}_k\boldsymbol{H}\boldsymbol{w}_k$ and
$\acute{\boldsymbol{\Psi}}_k = \boldsymbol{\Phi}_k\boldsymbol{H}\boldsymbol{w}_{\acute{k}}$ }.\\
The SOP constraint of \eqref{eq:9c} is derived from Theorem 1 and can be defined as
\begin{equation}
    S_k \ge \log_2{\Bigg(1 + \frac{\big|\boldsymbol{\theta}_k^H\boldsymbol{H}\boldsymbol{w}_k\big|^2\ln\delta^{-1}}{\big|\boldsymbol{\theta}_k^H\boldsymbol{H}\boldsymbol{w}_{\acute{k}}\big|^2 + \sigma_{e,k}^2}\Bigg)}, \forall k,\label{eq:10}
\end{equation}
where $\boldsymbol{\theta}_k = \text{diag}(\boldsymbol{\Phi}_k)$ and by using \eqref{eq:10}, optimization problem \eqref{P1} can be converted into 
\begin{subequations}
\label{P2}
\begin{align}
\textrm{ES:}\max_{\boldsymbol{w}_k, \boldsymbol{\theta}_k} 
 &  \quad \sum_{k}[\log_2{(1 + \gamma_{b,k})} - S_k]^+,\label{eq:11a}\\
\textrm{s.t.} \quad & \eqref{eq:9b},\eqref{eq:10},\eqref{eq:9d}-\eqref{eq:9g}.
\end{align}
\end{subequations}
It is evident from \eqref{eq:11a} that a decrease in $S_k$ would result in a higher value of the objective function. As a result, when \eqref{eq:10} equals, the optimal value of $S_k$ is found and can be represented as
\begin{equation}
    S_k^{*} = \log_2{\Bigg(1 + \frac{\big|\boldsymbol{\theta}_k^H\boldsymbol{H}\boldsymbol{w}_k\big|^2\ln\delta^{-1}}{\big|\boldsymbol{\theta}_k^H\boldsymbol{H}\boldsymbol{w}_{\acute{k}}\big|^2 + \sigma_{e,k}^2}\Bigg)}, \forall k.\label{eq:12}
\end{equation}
Upon inserting $S_k^{*}$ into \eqref{eq:11a}, the optimization problem \eqref{P2} is correspondingly transformed into
\begin{subequations}
\label{P3}
\begin{align}
\textrm{ES:}\max_{\boldsymbol{w}_k, \boldsymbol{\theta}_k} 
 &  \quad \sum_{k}[\log_2{(1 + \gamma_{b,k})} - S_k^{*}]^+,\label{eq:13a}\\
\textrm{s.t.} \quad & \eqref{eq:9b},\eqref{eq:9d}-\eqref{eq:9g}.
\end{align}
\end{subequations}
\subsection{Addressing Non-Convexity of Optimization Problem For ES}
To cope with the non-convex optimization problem of \eqref{P3}, we break down \eqref{P3} into manageable AO  approach, and we solve these subproblems iteratively\cite{setoode2024query,niu2021simultaneous,kavianinia2022resource}.

Initially, based on  \cite{niu2021weighted,sheng2018beamforming,kavianinia2023sum}, the subsequent two inequalities are valid around a given point $\Big\{\tilde{x},\tilde{y} \Big\}$:
\begin{equation}
    \begin{split}
    \ln{\Bigg(1+\frac{\big|x\big|^2}{y}\Bigg)} &\ge \ln{\Bigg(1+\frac{\big|\tilde{x}\big|^2}{\tilde{y}}\Bigg)} - \frac{\big|\tilde{x}\big|^2}{\tilde{y}}\\
        &+ \frac{2\mathfrak{R}\big\{x\tilde{x}\big\}}{\tilde{y}}
        -\frac{\big|\tilde{x}\big|^2\Big( y + \big|x\big|^2 \Big)}{\tilde{y}\Big( \tilde{y} + \big|\tilde{x}\big|^2 \Big)},\label{eq:14}
\end{split}
     \raisetag{50pt}
\end{equation}
\begin{equation}
    \ln{\Bigg(1 + \frac{x}{y}\Bigg)} \le \ln{\Bigg(1 + \frac{\tilde{x}}{\tilde{y}}\Bigg)} + \frac{\tilde{y}}{\tilde{x}+\tilde{y}}\Bigg(\frac{x}{y}-\frac{\tilde{x}}{\tilde{y}}\Bigg).\label{eq:15}
\end{equation}

Therefore, for a given point $\Big\{\boldsymbol{\tilde{\theta}}_k,\boldsymbol{\tilde{w}}_k \Big\}$, to handle the challenges posed by the non-convex objective function \eqref{eq:13a}, we obtain
\begin{equation}
\begin{split}
&\log_2{(1 + \gamma_{b,k})} \ge \frac{1}{\ln(2)}\bigg[ \log_2{\Big(1 + a_{1,k}\big|\boldsymbol{\tilde{\theta}}_k^H\boldsymbol{F}_k\boldsymbol{\tilde{w}}_k\big|^2\Big)}\\
& -a_{1,k}\big|\boldsymbol{\tilde{\theta}}_k^H\boldsymbol{F}_k\boldsymbol{\tilde{w}}_k\big|^2 + a_{1,k}2\mathfrak{R}\big\{\boldsymbol{w}_k^H\boldsymbol{F}_k^H\boldsymbol{\theta}_k\boldsymbol{\tilde{\theta}}_k^H\boldsymbol{F}_k\boldsymbol{\tilde{w}}_k \big\}\\
& - a_{2,k}\Big(\big|\boldsymbol{\theta}_k^H\boldsymbol{F}_k\boldsymbol{w}_k\big|^2 + \big|\boldsymbol{\theta}_k^H\boldsymbol{F}_k\boldsymbol{w}_{\acute{k}}\big|^2\Big)\bigg],
\end{split}
\raisetag{20pt}
\end{equation}
and
\begin{equation}
\begin{split}
S_k^* &\le \frac{1}{\ln(2)}\bigg[ \log_2{\Big(1 + a_{3,k}\Big) - \frac{a_{3,k}}{1 + a_{3,k}}}\\
&+ \frac{1}{1 + a_{3,k}}\frac{\big|\boldsymbol{\theta}_k^H\boldsymbol{H}\boldsymbol{w}_k\big|^2\ln\delta^{-1}}{2\mathfrak{R}\big\{\boldsymbol{w}_{\acute{k}}^H\boldsymbol{H}\boldsymbol{\theta}_k\boldsymbol{\tilde{\theta}}_k^H\boldsymbol{H}\boldsymbol{\tilde{w}}_{\acute{k}} \big\} + a_{4,k}}\bigg],\label{eq:17}
\end{split}
\raisetag{20pt}
\end{equation}
within the trust region, we have $2\mathfrak{R}\big\{\boldsymbol{w}_{\acute{k}}^H\boldsymbol{H}\boldsymbol{\theta}_k\boldsymbol{\tilde{\theta}}_k^H\boldsymbol{H}\boldsymbol{\tilde{w}}_{\acute{k}} \big\} - \big|\boldsymbol{\tilde{\theta}}_k^H\boldsymbol{H}\boldsymbol{\tilde{w}}_{\acute{k}}\big|^2 > 0$, and we apply $\big|\boldsymbol{\theta}_k^H\boldsymbol{H}\boldsymbol{w}_{\acute{k}}\big|^2 \ge 2\mathfrak{R}\big\{\boldsymbol{w}_{\acute{k}}^H\boldsymbol{H}\boldsymbol{\theta}_k\boldsymbol{\tilde{\theta}}_k^H\boldsymbol{H}\boldsymbol{\tilde{w}}_{\acute{k}} \big\} - \big|\boldsymbol{\tilde{\theta}}_k^H\boldsymbol{H}\boldsymbol{\tilde{w}}_{\acute{k}}\big|^2$ because $\big|\boldsymbol{\theta}_k^H\boldsymbol{H}\boldsymbol{w}_{\acute{k}}\big|^2$ is convex with respect to $\boldsymbol{\theta}_k$ and $\boldsymbol{w}_{\acute{k}}$. The constants $\{a_{1,k}, a_{2,k}, a_{3,k}, a_{4,k}\}$ are respectively defined by
\begin{subequations}
\label{eq:18}
\begin{align}
&a_{1,k} = \frac{1}{\big|\boldsymbol{\tilde{\theta}}_k^H\boldsymbol{F}_k\boldsymbol{\tilde{w}}_{\acute{k}}\big|^2 + \sigma_{b,k}^2},\\
a_{2,k} = &\frac{a_{1,k}\big|\boldsymbol{\tilde{\theta}}_k^H\boldsymbol{F}_k\boldsymbol{\tilde{w}}_k\big|^2}{\big|\boldsymbol{\tilde{\theta}}_k^H\boldsymbol{F}_k\boldsymbol{\tilde{w}}_k\big|^2 + \big|\boldsymbol{\tilde{\theta}}_k^H\boldsymbol{F}_k\boldsymbol{\tilde{w}}_{\acute{k}}\big|^2 + \sigma_{b,k}^2},\\
&a_{3,k} =  \frac{\big|\boldsymbol{\tilde{\theta}}_k^H\boldsymbol{H}\boldsymbol{\tilde{w}}_{k}\big|^2}{\big|\boldsymbol{\tilde{\theta}}_k^H\boldsymbol{H}\boldsymbol{\tilde{w}}_{\acute{k}}\big|^2 + \sigma_{e,k}^2},\\
&a_{4,k}= -\big|\boldsymbol{\tilde{\theta}}_k^H\boldsymbol{H}\boldsymbol{\tilde{w}}_{\acute{k}}\big|^2 + \sigma_{e,k}^2.
\end{align}
\end{subequations}

In addition, the energy harvesting constraint \eqref{eq:9d} is non-convex, based on \eqref{app:3} and \eqref{app:4}, the harvested energy at $\text{Eve}_k$ is computed as
\begin{equation}
    E_{e,k} =  \bigg(\big| \boldsymbol{\theta}_k^H \boldsymbol{H} \boldsymbol{w}_k\big|^2 + \big| \boldsymbol{\theta}_k^H \boldsymbol{H}\boldsymbol{w}_{\acute{k}}\big|^2\bigg).\label{eq:19}
\end{equation}
Firstly, we can reformulate the constraint \eqref{eq:9d} as  
\begin{equation}
    E_{e,k} \ge E_{\text{min}} \longrightarrow \log_2{(1 + E_{e,k})} \ge \log_2{(1 + E_{\text{min}})},\label{eq:20}
\end{equation}
secondly, after inserting \eqref{eq:19} into \eqref{eq:20}, the upper bound of the left-hand side of \eqref{eq:20} is determined by applying \eqref{eq:15} as
\begin{equation}
\begin{split}
&\log_2{(1 + E_{e,k})} \le\\
&\frac{1}{\ln(2)}\Bigg[ \log_2{\Big(1 + \big| \boldsymbol{\tilde{\theta}}_k^H \boldsymbol{H} \boldsymbol{\tilde{w}}_k\big|^2
+ \big| \boldsymbol{\tilde{\theta}}_k^H \boldsymbol{H}\boldsymbol{\tilde{w}}_{\acute{k}}\big|^2\Big)} \\
&+ \frac{1}{\big| \boldsymbol{\tilde{\theta}}_k^H \boldsymbol{H} \boldsymbol{\tilde{w}}_k\big|^2 + \big| \boldsymbol{\tilde{\theta}}_k^H \boldsymbol{H}\boldsymbol{\tilde{w}}_{\acute{k}}\big|^2}
\Bigg(\big| \boldsymbol{\theta}_k^H \boldsymbol{H} \boldsymbol{w}_k\big|^2 \\
&+ \big| \boldsymbol{\theta}_k^H \boldsymbol{H}\boldsymbol{w}_{\acute{k}}\big|^2 -\big| \boldsymbol{\tilde{\theta}}_k^H \boldsymbol{H} \boldsymbol{\tilde{w}}_k\big|^2 - \big| \boldsymbol{\tilde{\theta}}_k^H \boldsymbol{H}\boldsymbol{\tilde{w}}_{\acute{k}}\big|^2 \Bigg)\Bigg] = E_{e,k}^*,\label{eq:21}
\end{split}
\raisetag{40pt}
\end{equation}
thus, the reformulated constraint is given by
\begin{equation}
    E_{e,k}^* \ge \log_2{(1 + E_{\text{min}})}, \label{eq:22}
\end{equation}
which is convex constraint.

Using the aforementioned equations and ignoring the constant terms, \eqref{P3} can be reformulated around the provided point $\Big\{\boldsymbol{\tilde{\theta}}_k,\boldsymbol{\tilde{w}}_k \Big\}$ as
\begin{subequations}
\label{P4}
\begin{align}
\mathcal{P}_{1}:\min _{\boldsymbol{\theta}_k,\boldsymbol{w}_k} & 
\begin{aligned}
\raisetag{60pt}
&\sum_k\left\{- a_{1, k} 2 \Re\left\{\boldsymbol{w}_k^H \boldsymbol{F}_k^H \boldsymbol{\theta}_k \tilde{\boldsymbol{\theta}}_k^H \boldsymbol{F}_k \tilde{\boldsymbol{w}}_k\right\}\right. \\
& + a_{2, l}\left(\left|\boldsymbol{\theta}_k^H \boldsymbol{F}_k \boldsymbol{w}_k\right|^2+\left|\boldsymbol{\theta}_k^H \boldsymbol{F}_k \boldsymbol{w}_{k^{\prime}}\right|^2\right) \\
& \left.+\frac{1}{1+a_{3, k}} \frac{\left|\boldsymbol{\theta}_k^H \boldsymbol{H}_k \boldsymbol{w}_k\right|^2\ln\delta^{-1}}{2 \Re\left\{\boldsymbol{w}_{k^{\prime}}^H \boldsymbol{H}_k^H \boldsymbol{\theta}_k \tilde{\boldsymbol{\theta}}_k^H \boldsymbol{H}_k \tilde{\boldsymbol{w}}_{k^{\prime}}\right\}+a_{4, k}}\right\} \label{eq:23a}
\end{aligned}\\
\text { s.t. } & \quad \eqref{eq:9b},\eqref{eq:22},\label{eq:23b} \\
& \quad \operatorname{diag}\left(\sum_l \boldsymbol{\theta}_l \boldsymbol{\theta}_l^H\right)=\mathbf{1}.\label{eq:23c}
\end{align}
\end{subequations}

According to \cite{boyd2004convex}, the objective function \eqref{eq:23a} has three distinct types of functions: the first term is a linear function of either $\boldsymbol{\theta}_k$ or $\boldsymbol{w}_k$; the second term is a quadratic function of either $\boldsymbol{\theta}_k$ or $\boldsymbol{w}_k$; and the third term is a convex quadratic-over-linear function of either $\boldsymbol{\theta}_k$ or $\boldsymbol{w}_k$. As a result, the optimization problem \eqref{P4} is convex with respect to $\boldsymbol{w}_k$ when $\boldsymbol{\theta}_k$ is fixed and can be solved using well-known  optimization toolbox.

Our next main priority is to optimize $\boldsymbol{\theta}_k$. Although \eqref{eq:23a} is convex with respect to $\boldsymbol{\theta}_k$ when $\tilde{\boldsymbol{w}}_k$ is provided, \eqref{eq:23c} presents a significant challenge, causing \eqref{P4} to be non-convex. To address this, we first linearize \eqref{eq:23c} and solve for $\varkappa_{k,m}$ which is defined as $\varkappa_{k,m} = {[\boldsymbol{\theta}_k]_m}^{*} [\boldsymbol{\theta}_k]_m, \forall m \in \mathcal{M}$, by introducing auxiliary vectors $\boldsymbol{\varkappa}_{k} = [\varkappa_{k,1},\dots, \varkappa_{k,M}]^T$. We then relax the inequality $\varkappa_{k,m} = {[\boldsymbol{\theta}_k]_m}^{*} [\boldsymbol{\theta}_k]_m$ by allowing $\varkappa_{k,m}\le {[\boldsymbol{\theta}_k]_m}^{*} [\boldsymbol{\theta}_k]_m \le\varkappa_{k,m}$ with respect to PCCP approach. Furthermore, $\varkappa_{k,m} \le 2\Re\{{[\boldsymbol{\theta}_k]_m}^{*} [\boldsymbol{\tilde{\theta}}_k]_m\} - {[\boldsymbol{\tilde{\theta}}_k]_m}^{*} [\boldsymbol{\tilde{\theta}}_k]_m $ can be used to approximate $\varkappa_{k,m}\le {[\boldsymbol{\theta}_k]_m}^{*} [\boldsymbol{\theta}_k]_m \le\varkappa_{k,m}$ \cite{zhou2020framework}.
\begin{figure*}
\begin{subequations}
\label{P5}
\begin{align}
 \mathcal{P}_{2}:&\min _{\boldsymbol{\theta}_k, \boldsymbol{\varkappa}_{k}, \boldsymbol{\varsigma}_{k}}
 \begin{aligned}
 \label{eq:24a}
 &\sum_k\left\{- a_{1, k} 2 \Re\left\{\tilde{\boldsymbol{w}}_k^H \boldsymbol{F}_k^H \boldsymbol{\theta}_k \tilde{\boldsymbol{\theta}}_k^H \boldsymbol{F}_k \tilde{\boldsymbol{w}}_k\right\}+ a_{2, k}\left(\left|\boldsymbol{\theta}_k^H \boldsymbol{F}_k \tilde{\boldsymbol{w}}_k\right|^2+\left|\boldsymbol{\theta}_k^H \boldsymbol{F}_k \tilde{\boldsymbol{w}}_{k^{\prime}}\right|^2\right)\right.\\
 &    \left.+\frac{1}{1+a_{3, k}} \frac{\left|\boldsymbol{\theta}_k^H \boldsymbol{H}_k \tilde{\boldsymbol{w}}_k\right|^2\ln\delta^{-1}}{2 \Re\left\{\tilde{\boldsymbol{w}}_{k^{\prime}}^H \boldsymbol{H}_k^H \boldsymbol{\theta}_k \tilde{\boldsymbol{\theta}}_k^H \boldsymbol{H}_k \tilde{\boldsymbol{w}}_{k^{\prime}}\right\}+a_{4, k}}\right\}+\lambda^{[i]} \sum_k \sum_{m=1}^{2 M} \varsigma_{k,m} 
 \end{aligned}\\
& \quad \text { s.t. }  \quad 
\eqref{eq:22},\\
&\quad \quad \quad \quad   \left[\boldsymbol{\theta}_k\right]_m^*\left[\boldsymbol{\theta}_k\right]_m \leq \varkappa_{k,m}+\varsigma_{k,m},\left[\tilde{\boldsymbol{\theta}}_k\right]_m^*\left[\tilde{\boldsymbol{\theta}}_k\right]_m-2 \Re\left\{\left[\boldsymbol{\theta}_k\right]_m^*\left[\tilde{\boldsymbol{\theta}}_k\right]_m\right\} \leq \varsigma_{k, m+M}-\varkappa_{l, m}, \forall m \in \mathcal{M} \text {, } \\
& \quad \quad \quad \quad \varkappa_{r, m}+\varkappa_{t, m}=1, \varkappa_{r, m} \geq 0, \varkappa_{t, m} \geq 0, \forall m \in \mathcal{M} \text {, }
\end{align}
\end{subequations}
\hrulefill
\end{figure*}

Therefore, we derive the algorithm, which is illustrated at the top of the next page. The slack variable for the modulus constraints is denoted by $\varsigma_{k,m} \ge 0$, and the penalty term added to the objective function is represented by $\sum_{m=1}^{2M}\varsigma_{k,m}$. This term is scaled by the multiplier $\lambda^{[i]}$ in the $i$-th iteration. Furthermore, to update $\lambda^{[i]}$,  $\lambda^{[i]} = \min\{\beta\lambda^{[i-1]},\lambda_{\text{max}}\}$ is utilized, where the upper bound $\lambda_{\text{max}}$ is utilized to prevent numerical problems.
The recommended alternative method for resolving the \eqref{P3} sub-problems is summed up in \textbf{Algorithm 1} based on the previous analysis.
\subsection{Addressing Non-Convexity of Optimization Problem For MS}
It has been recognized that by substituting $\alpha_m^k \in [0,1]$ with $\alpha_m^k \in \{0,1\}$, we can reformulate \eqref{P1} to the MS protocol optimization problem. The optimization of $\boldsymbol{\theta}_k$ necessitates addressing a non-convex mixed-integer problem, which is resolved by introducing the slack variable $q_{k,m}$. It is confirmed that $\varkappa_{k,m} \in \{0, 1\}$ is equal to $\varkappa_{k,m} = q_{k,m}$ and $\varkappa_{k,m}(1 - q_{k,m}) = 0$ \cite{hua2021uav}. Subsequently, the penalty term $\kappa^{[i]}\sum_k\sum_{m=1}^{M}(\big|q_{k,m} - \varkappa_{k,m}\big|^2 + \big|q_{k,m}(1 - \varkappa_{k,m})\big|^2)$ is incorporated into the objective function \eqref{eq:24a}, where $\kappa^{[i]}$ serves as the penalty factor in the $i$-th iteration and is updated in the same manner as $\lambda^{[i]}$.\\
Given $\{\boldsymbol{\theta}_k, \boldsymbol{\varkappa}_k, \boldsymbol{\varsigma}_k\}$, the optimal $q_{k,m}$ can be obtained through first-order optimality, such as \cite{hua2021uav}
\begin{equation}
    q_{k,m} = \frac{\varkappa_{k,m} + (\varkappa_{k,m})^2}{1 + (\varkappa_{k,m})^2}.\label{eq:25}
\end{equation}
Conversely, with a given $q_{k,m}$ the remaining variables can be solved using the previously proposed PCCP method. \textbf{Algorithm 2 } summarizes the proposed alternate approach for handling the generalized \eqref{P3} sub-problems for MS protocol. 

\begin{algorithm}[t!]
\caption{An iterative method for the ES protocol to address the problem outlined in \eqref{P3}.}\label{alg:cap1}
\textbf{Input}: Initial values for $\boldsymbol{\tilde{\theta}}_{k}^{(1)}$ and $\boldsymbol{\tilde{w}}_{k}^{(1)}, \forall k$, Channel coefficients $\boldsymbol{H}$, $\boldsymbol{h}_{k}$, and $\boldsymbol{v}_{k}, \forall k$. Maximum power $P_{\text{max}}$, SOP upper bound $\delta$, minimum required energy $E_{\text{min}}$, $\lambda^{[1]}, \lambda_{\text{max}}, \beta$ and tolerance $\epsilon$. 
\begin{algorithmic}[1]
\For{$i=1, 2, \dots$}
    \State
    \parbox[t]{\dimexpr\linewidth-\algorithmicindent}{%
    For given $\boldsymbol{\tilde{\theta}}_{k}^{(i)}$ and $\boldsymbol{\tilde{w}}_{k}^{(i)}, \forall k,   $ update $\boldsymbol{\theta}_{k}^{(i)}, \forall k$ using $\mathcal{P}_{2}$.
    }
    \State
    \parbox[t]{\dimexpr\linewidth-\algorithmicindent}{%
    For given $\boldsymbol{\tilde{\theta}}_{k}^{(i)}$,  $\boldsymbol{\tilde{w}}_{k}^{(i)}$, and $\boldsymbol{\theta}_{k}^{(i)}, \forall k$  update $\boldsymbol{w}_{k}^{(i)}, \forall k$ using $\mathcal{P}_{1}$.
    }
    \State
    \parbox[t]{\dimexpr\linewidth-\algorithmicindent}{%
    Update $\boldsymbol{\tilde{\theta}}_{k}^{(i+1)} = \boldsymbol{\theta}_{k}^{(i)}$ and $\boldsymbol{\tilde{w}}_{k}^{(i+1)} = \boldsymbol{w}_{k}^{(i)}$.
    }
    \State
    \parbox[t]{\dimexpr\linewidth-\algorithmicindent}{%
    Update $\lambda^{[i+1]} = \min\{\beta\lambda^{[i]},\lambda_{\text{max}}\}$.
    }
    
    \State \textbf{Until} $\Big|{\textrm{SSR}}^{(i+1)} - {\textrm{SSR}}^{(i)}  \Big| < \epsilon$.
\EndFor
\end{algorithmic}
\textbf{Output}: The optimal solutions: ${\boldsymbol{\mathit{w}}_k}^{\textrm{opt}}={\boldsymbol{\mathit{w}}_k}^{(i)}$ and $\boldsymbol{\theta}_k^{\text{opt}}= \boldsymbol{\theta}_k^{(i)}, \forall k$.
\end{algorithm}
\begin{algorithm}[ht!]
\caption{An iterative method for the MS protocol to address the generalized problem outlined in \eqref{P3}.}\label{alg:cap2}
\textbf{Input}: Initial values for $\boldsymbol{\tilde{\theta}}_{k}^{(1)}$, $\boldsymbol{\tilde{w}}_{k}^{(1)}$ and $\boldsymbol{q}_k, \forall k$, Channel coefficients $\boldsymbol{H}$, $\boldsymbol{h}_{k}$, and $\boldsymbol{v}_{k}, \forall k$. Maximum power $P_{\text{max}}$, SOP upper bound $\delta$, minimum required energy $E_{\text{min}}$, $\lambda^{[1]}, \lambda_{\text{max}}, \kappa^{[1]}, \kappa_{\text{max}}, \beta$ and tolerance $\epsilon$. 
\begin{algorithmic}[1]
\For{$i=1, 2, \dots$}
    \State
    \parbox[t]{\dimexpr\linewidth-\algorithmicindent}{%
    For given $\boldsymbol{\tilde{\theta}}_{k}^{(i)}$ and $\boldsymbol{\tilde{w}}_{k}^{(i)}, \forall k,   $ update $\boldsymbol{\theta}_{k}^{(i)}, \forall k$ using generalized $\mathcal{P}_{2}$.
    }
    \State
    \parbox[t]{\dimexpr\linewidth-\algorithmicindent}{%
    For given $\boldsymbol{\tilde{\theta}}_{k}^{(i)}$,  $\boldsymbol{\tilde{w}}_{k}^{(i)}$, and $\boldsymbol{\theta}_{k}^{(i)}, \forall k$  update $\boldsymbol{w}_{k}^{(i)}, \forall k$ using  $\mathcal{P}_{1}$.
    }
    \State
    \parbox[t]{\dimexpr\linewidth-\algorithmicindent}{%
    For given $\{\boldsymbol{\theta}_k, \boldsymbol{\varkappa}_k, \boldsymbol{\varsigma}_k\}$  update $\boldsymbol{q}_{k}^{(i +1)}, \forall k$ using \eqref{eq:25}.
    }
    \State
    \parbox[t]{\dimexpr\linewidth-\algorithmicindent}{%
    Update $\boldsymbol{\tilde{\theta}}_{k}^{(i+1)} = \boldsymbol{\theta}_{k}^{(i)}$ and $\boldsymbol{\tilde{w}}_{k}^{(i+1)} = \boldsymbol{w}_{k}^{(i)}$.
    }
    \State
    \parbox[t]{\dimexpr\linewidth-\algorithmicindent}{%
    Update $\lambda^{[i+1]} = \min\{\beta\lambda^{[i]},\lambda_{\text{max}}\}$.
    }
    \State
    \parbox[t]{\dimexpr\linewidth-\algorithmicindent}{%
    Update $\kappa^{[i+1]} = \min\{\beta\kappa^{[i]},\kappa_{\text{max}}\}$.
    }
    
    \State \textbf{Until} $\Big|{\textrm{SSR}}^{(i+1)} - {\textrm{SSR}}^{(i)}  \Big| < \epsilon$.
\EndFor
\end{algorithmic}
\textbf{Output}: The optimal solutions: ${\boldsymbol{\mathit{w}}_k}^{\textrm{opt}}={\boldsymbol{\mathit{w}}_k}^{(i)}$ and $\boldsymbol{\theta}_k^{\text{opt}}= \boldsymbol{\theta}_k^{(i)}, \forall k$.
\end{algorithm}
\subsection{Complexity Analysis}
The computational complexity of the proposed design is detailed in this section, following the methodology outlined in \cite{wang2014outage}. Specifically, the general expression for the ES or MS scheme is $\ln{(\frac{1}{\epsilon})}\sqrt{\eta}\vartheta$. The values of $\eta$ and $\vartheta$, which are determined by the number and dimension of constraints in the sub-problems, remain constant. Meanwhile, $\epsilon$ represents the desired level of accuracy. Thus,
the time complexity for beamforming and TaRCs optimization sub-problem in both ES and MS protocol are computed as $\mathcal{O}\big({N_t}^2\big)$ and $\mathcal{O}\big({M}^2\big)$, respectively.
\section{Numerical Results and Discussion}
In order to evaluate the performance of the proposed secure transmission algorithm, simulations are conducted. The locations and channels of both Eves and the independent Bobs are documented for each simulation experiment, resulting in an average of over 100 simulation trials. Moreover, the maximum number of iterations and stopping accuracy for each algorithm are set at 30 and $10^{-3}$, respectively. We assume that the distances between the STAR-RS and Bobs, as well as the distance between the BS and the STAR-RIS, are set at $d =10$ m, for the simulation scenario. A randomly generated distance of 0 to 10 meters separates Eves from the STAR-RIS. In addition, the path loss is calculated by using the formula $P_L = PL_{0}(\frac{d}{d_0})^{-\alpha}$. Here, $PL_0$ is the channel gain at a reference distance $d_0$ of 1 meter, $d$ is the path distance, and $\alpha$ is the path loss exponent. Here are the parameters for the simulation: $P_{\text{max}} = 0$ dBW, $N_t = 4$, $M = 40$, $\sigma_{b,k}^2 = \sigma_{e,k}^2 = 1$, $\alpha = 2.2$. Additionally, we set $\lambda^{[1]} = \kappa^{[1]} = 20$, $\lambda_{\text{max}} = \kappa_{\text{max}} = 10$, and $\beta = 0.8$ as the PCCP approach's parameters. The parameter values for large-scale fading are considered to be constant and already known for the duration of the simulations. Each element of $\boldsymbol{H}$ is chosen from $\mathcal{C}\mathcal{N} (0,1)$, and the small-scale fading vectors from the STAR-RIS to the Bobs and all Eves are generated individually according to $\mathcal{C}\mathcal{N} (\bold{0},\boldsymbol{I}_m)$.
In order to demonstrate the benefits of the proposed scheme in terms of secure transmission and fulfilling energy requirements at Eves, we first compare it to a conventional RIS while accounting for the imperfect CSI of Eves' channels. Furthermore, by comparing the proposed approach for the ES and MS protocols with STAR-RIS-assisted secure transmission which ignores CSI uncertainty, we show the need of accounting for imperfect CSI for Eves\cite{li2020massive}.

In this case, we evaluate the proposed method with various baselines: 1) The ES protocol with imperfect CSI (IPCSI); 2) The MS protocol with IPCSI; 3) The conventional RIS with IPCSI by employing RISs involves positioning a transmitting-only RIS and a reflecting-only RIS side by side at the same location as the STAR-RIS, with each RIS comprised of $M/2$ elements \cite{mu2021simultaneously}; 4) The ES protocol with perfect CSI (PCSI); 5) The MS protocol with PCSI.

\begin{figure}[t!]
\centering
\includegraphics[width=0.45\textwidth]{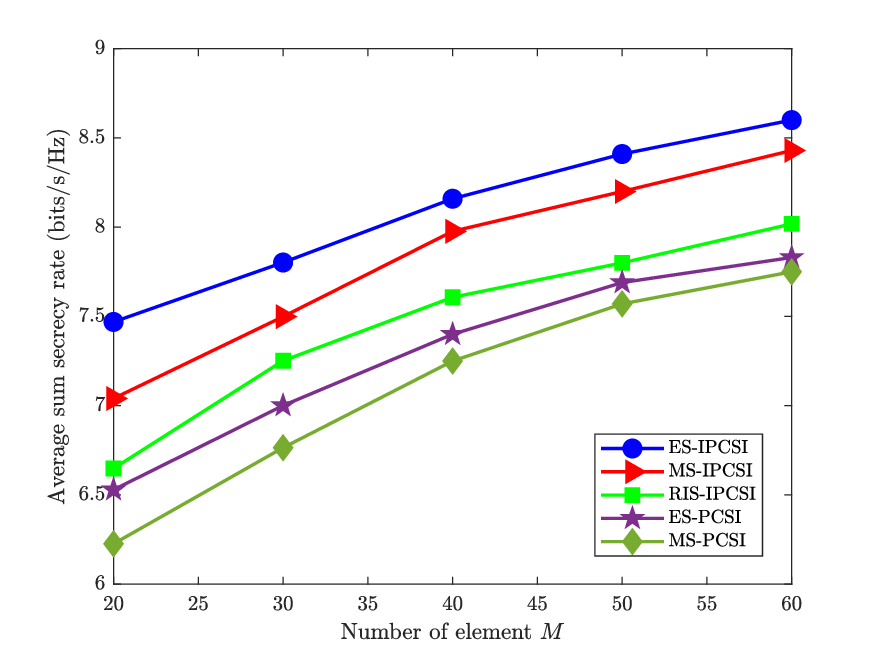}
    \caption{Average sum secrecy rate versus STAR-RIS number of elements: $P_{\text{max}} = 0$ dBW, $E_{\text{min}}= -20$ dB, $N_t = 4$, $\delta = 0.5$.}
    \label{fig:2}
    \vspace{-.3 cm}
\end{figure}

\begin{figure}[t]
    \centering
\includegraphics[width=0.45\textwidth]{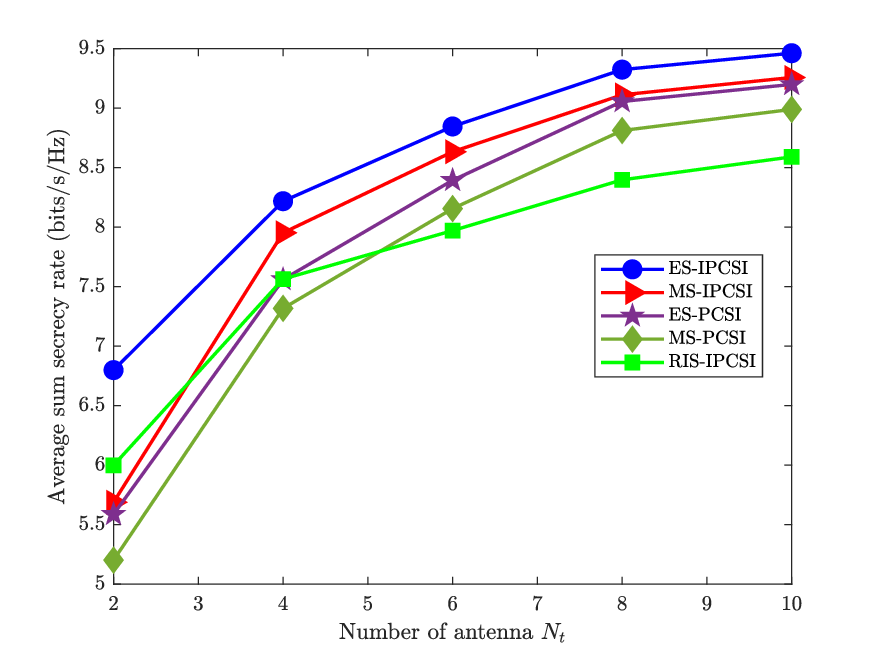}
    \caption{Average sum secrecy rate versus number of antennas: $P_{\text{max}} = 0$ dBW, $E_{\text{min}}= -20$ dB, $M = 40$, $\delta = 0.5$.}
    \label{fig:3}
    \vspace{-.3 cm}
\end{figure}

\begin{figure}[t!]
    \centering
\includegraphics[width=0.45\textwidth]{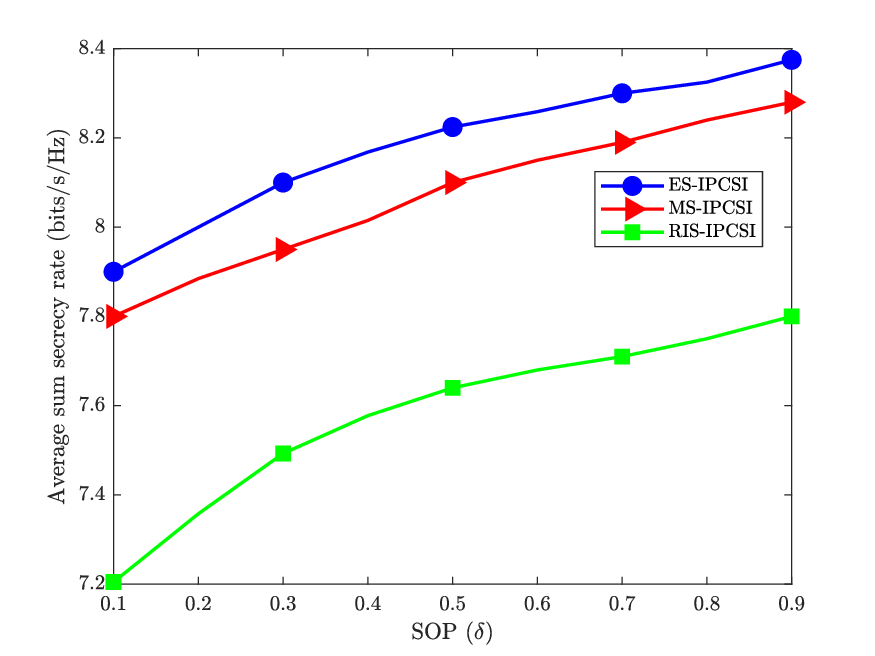}
    \caption{Average sum secrecy rate versus upper bound SOP: $P_{\text{max}}= 0$ dBW, $E_{\text{min}}= -20$ dB, $N_t = 4$, $M = 40$.}
    \label{fig:6}
    \vspace{-.3 cm}
\end{figure}

First, we demonstrate in Fig. \ref{fig:2} how raising the STAR-RIS elements impacts secure transmission. This figure illustrates the enhanced sum secrecy rate as the number of elements $M$ grows, while fulfilling the energy requirements at Eves. The "ES-IPCSI" demonstrates consistent superiority, closely followed by "MS-IPCSI". Meanwhile, "RIS-IPCSI" demonstrates slightly better performance than "ES-PCSI" in intermediate levels, and "MS-PCSI" exhibits the least performance among the five methods. Then, in Fig. \ref{fig:3}, average sum secrecy rate versus number of antennas is depicted. The sum secrecy rate improves with respect to $N_t$ for all baselines, since increasing the number of antennas incorporates diversity to secure communication. Furthermore, the ES protocol exhibits superior performance compared to the MS protocol, due to the optimized components of the STAR-RIS system for signal reflection and transmission in the ES protocol. On the other hand, the MS protocol only enables the STAR-RIS system to transmit or reflect received signals. It's worth noting that the ES and MS-based STAR-RIS consistently surpasses the performance of conventional RIS with IPCSI. Also, it can be seen that all baselines with IPCSI show better performance   
\begin{figure}[ht]
    \centering
\includegraphics[width=0.45\textwidth]{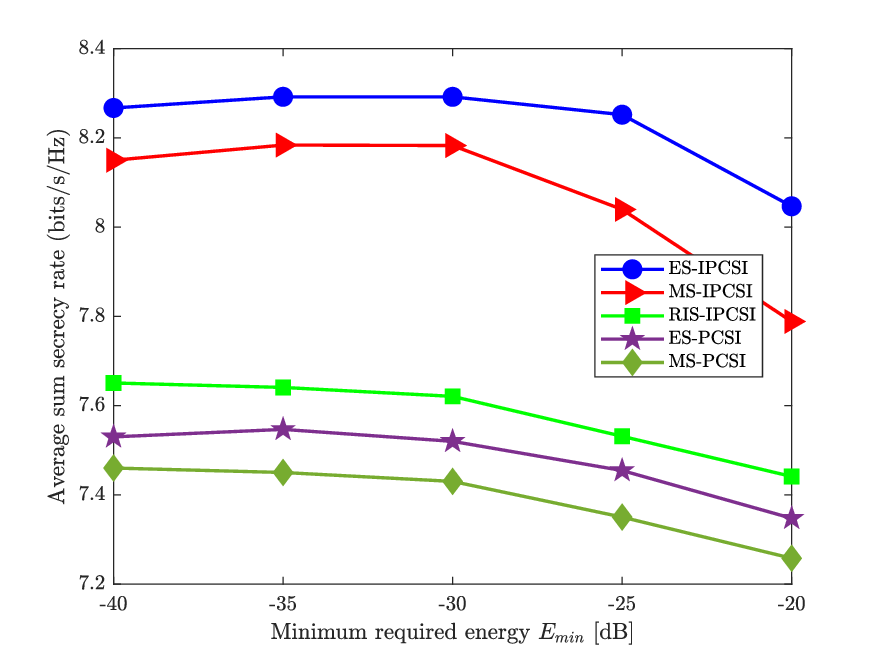}
    \caption{Average sum secrecy rate versus min required energy: $P_{\text{max}} = 0$ dBW, $N_t = 4$ dB, $M = 40$, $\delta = 0.5$.}
    \label{fig:4}
    \vspace{-.3 cm}
\end{figure}
than PCSI. In Fig. \ref{fig:4}, the sum secrecy rate is illustrated to be a decreasing function of Eves' minimal energy demand. In addition, raising the minimum demanded energy at Eves requires STAR-RIS elements to be committed to Eve's channel, hence meeting harvested energy constraints that are inconsistent with secure transmission. Also, the STAR-RIS protocols with IPCSI outperform convectional RIS with PCSI and STAR-RIS with PCSI consideration. As a result, the trade-off between improved secure transmission and energy charging must be managed.
\begin{figure}[t!]
    \centering
\includegraphics[width=0.45\textwidth]{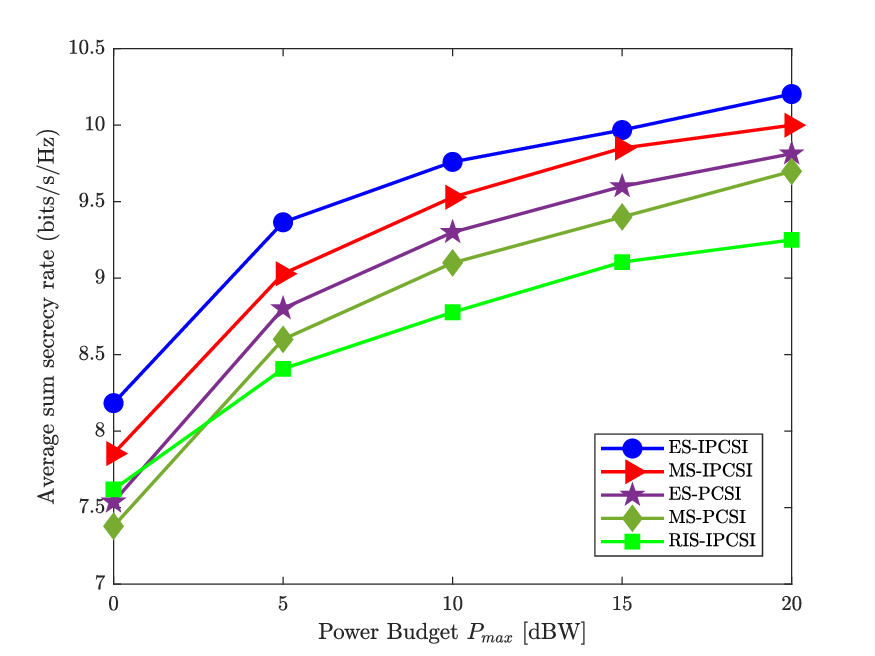}
    \caption{Average sum secrecy rate versus power budget: $E_{\text{min}}= -20$ dB, $N_t = 4$, $M = 40$, $\delta = 0.5$.}
    \label{fig:5}
    \vspace{-.3 cm}
\end{figure}
Fig. \ref{fig:5} demonstrates that the proposed STAR-RIS system greatly improves average sum secrecy rates as $P_{\text{max}}$ values increase. This fact suggests that the proposed approach may fully capitalize on transmit power to increase secure transmission performance while charging Eves' batteries. Furthermore, when $P_{\text{max}}$ increases, the performance disparities between the proposed solution and alternative baseline schemes widen, demonstrating the strength of the proposed approach and IPCSI's superiority over PCSI. Fig. \ref{fig:6} shows how the SOP limitation affects secure transmission performance, as all average sum secrecy rates increase. The proposed technique for ES and MS protocols consistently outperforms conventional RIS with respect to IPCSI. In addition, for sake of STAR-RIS protocols comparsion, the ES protocol performs better with the same settings because STAR-RIS components can be configured to reflect and transmit signals optimally. Whereas, they can only use the MS protocol to transmit or reflect received signals.
\section{Conclusions}
In this paper, We evaluated the effectiveness and potential of a novel STAR-RIS in improving secure communication within a MISO wiretap network. Our study included various scenarios, such as energy harvesting by eavesdroppers and the impact of imperfect CSI on secure transmission. To maximize the sum secrecy rate and meet the energy harvested by the eavesdroppers, we optimized the transmit Beamformer, STAR-RIS TaRCs, and transmission rate using the PCCP algorithm based on alternating optimization. Our findings demonstrate that a well-optimized STAR-RIS significantly outperforms conventional RIS, particularly in situations with probabilistic constraints and imperfect CSI, enhancing the average sum secrecy rate while fulfilling energy constraints. Simulation outcomes indicate that in high-power domains with a substantial number of antennas and a significant number of STAR-RIS elements, the ES and MS protocols are particularly effective in scenarios with imperfect CSI compared to perfect CSI, even STAR-RIS outperforms convectional RIS in each scenario. The study supports the notion that STAR-RIS technology provides substantial advantages in secure communication with energy-harvesting eavesdroppers, making it a promising solution for future wireless networks.

\section*{APPENDIX A\\Proof of Theorem 1}
\addcontentsline{toc}{section}{APPENDIX A: Proof of Theorem 1}
Using the notation $\boldsymbol{\Psi}_k = \boldsymbol{\Phi}_k\boldsymbol{H}\boldsymbol{w}_k \in \mathbb{C}^{M \times 1}$ and $\acute{\boldsymbol{\Psi}}_k = \boldsymbol{\Phi}_k\boldsymbol{H}\boldsymbol{w}_{\acute{k}} \in \mathbb{C}^{M \times 1}$, the SOP of \eqref{eq:6} can be defined as
\begin{equation}
    \begin{split}
       p_{so}^k&= \text{Pr}\Bigg\{\log_2{\Bigg( 1 + \frac{\big| \boldsymbol{v}_k^H \boldsymbol{\Psi}_k\big|^2}{\big| \boldsymbol{v}_k^H \acute{\boldsymbol{\Psi}}_k\big|^2 + \sigma_{e,k}^2} \Bigg)} > S_k \Bigg\}\\
       &= \text{Pr}\Bigg\{\big|\boldsymbol{v}_k^H \boldsymbol{\Psi}_k\big|^2 > (\big|\boldsymbol{v}_k^H \acute{\boldsymbol{\Psi}}_k\big|^2 + \sigma_{e,k}^2)(2^{S_k} -1) \Bigg\}\\
       &= \text{Pr}\Bigg\{\boldsymbol{\Psi}_k^H\boldsymbol{v}_k\boldsymbol{v}_k^H \boldsymbol{\Psi}_k > (\acute{\boldsymbol{\Psi}}_k^H\boldsymbol{v}_k\boldsymbol{v}_k^H \acute{\boldsymbol{\Psi}}_k + \sigma_{e,k}^2)(2^{S_k} - 1) \Bigg\}.\label{eq:app1}
    \end{split}
     \raisetag{50pt}
 \end{equation}
Given the known value of $\boldsymbol{v}_k$ with respect to $ \boldsymbol{v}_k \sim \mathcal{C}\mathcal{N} (\bold{0},\boldsymbol{I}_m)$, the expected value of the random variables $\boldsymbol{\Psi}_k^H\boldsymbol{v}_k\boldsymbol{v}_k^H \boldsymbol{\Psi}_k$ and $\acute{\boldsymbol{\Psi}}_k^H\boldsymbol{v}_k\boldsymbol{v}_k^H \acute{\boldsymbol{\Psi}}_k$ can be obtained, respectively via \cite{kandukuri2002optimal}
\begin{equation}
    \mathbb{E} \Big\{\boldsymbol{\Psi}_k^H\boldsymbol{v}_k\boldsymbol{v}_k^H \boldsymbol{\Psi}_k\Big\} = \boldsymbol{\Psi}_k^H\boldsymbol{\Psi}_k,\label{app:3}
\end{equation}
\begin{equation}
    \mathbb{E} \Big\{\acute{\boldsymbol{\Psi}}_k^H\boldsymbol{v}_k\boldsymbol{v}_k^H \acute{\boldsymbol{\Psi}}_k\Big\} = \acute{\boldsymbol{\Psi}}_k^H\acute{\boldsymbol{\Psi}}_k.\label{app:4}
\end{equation}
Furthermore, note that the received signal power follows an exponential distribution\cite{stuber2001principles}. In light of this, we get $\boldsymbol{\Psi}_k^H\boldsymbol{v}_k\boldsymbol{v}_k^H \boldsymbol{\Psi}_k \sim \exp{(\boldsymbol{\Psi}_k^H\boldsymbol{\Psi}_k)}$ and $\acute{\boldsymbol{\Psi}}_k^H\boldsymbol{v}_k\boldsymbol{v}_k^H \acute{\boldsymbol{\Psi}}_k \sim \exp{(\acute{\boldsymbol{\Psi}}_k^H\acute{\boldsymbol{\Psi}}_k)}$, thus \eqref{eq:app1} can be further rewritten as
\begin{equation}
    \begin{split}
    &\text{Pr}\Bigg\{\boldsymbol{\Psi}_k^H\boldsymbol{v}_k\boldsymbol{v}_k^H \boldsymbol{\Psi}_k > (\acute{\boldsymbol{\Psi}}_k^H\boldsymbol{v}_k\boldsymbol{v}_k^H \acute{\boldsymbol{\Psi}}_k + \sigma_{e,k}^2)(2^{S_k} -1)\Bigg\} \\
    &=\exp{\Bigg(-\frac{(\acute{\boldsymbol{\Psi}}_k^H\acute{\boldsymbol{\Psi}}_k + \sigma_{e,k}^2)(2^{S_k} -1)}{\boldsymbol{\Psi}_k^H\boldsymbol{\Psi}_k}}\Bigg).
    \end{split}\label{eq:app2}
 \end{equation}  
A closed-form expression of $p_{so}^k$ can be found by substituting $\boldsymbol{\Psi}_k = \boldsymbol{\Phi}_k\boldsymbol{H}\boldsymbol{w}_k $ and $\acute{\boldsymbol{\Psi}}_k = \boldsymbol{\Phi}_k\boldsymbol{H}\boldsymbol{w}_{\acute{k}}$ into equation \eqref{eq:app2}.

%%====> References <===%%
\bibliographystyle{IEEEtran}
\bibliography{Secure_STAR_RIS}

\end{document}